\documentclass[aps,prd,nofootinbib]{revtex4-1}

\usepackage{amsmath,amssymb,amsfonts}
\usepackage{graphicx}
\usepackage{hyperref}
\usepackage{dashbox}
\usepackage{physics}

\begin{document}

\title{Beyond the Standard Model of Cosmology: Testing new paradigms with a \\ Multiprobe Exploration of the Dark Universe}
\author{Juan Garc\'ia-Bellido}
\affiliation{Instituto de F\'isica Te\'orica UAM/CSIC, Universidad Aut\'onoma de Madrid, 28049 Madrid, Spain}
\date{\today}

\begin{abstract}
Cosmology is living through fascinating times, where new observations from ground and space telescopes are questioning the established paradigm, the so-called $\Lambda$ Cold Dark Matter model. The particle nature of Dark Matter is severely constrained by underground experiments, while recent observations by galaxy surveys indicate that the cosmological constant ($\Lambda$) may not be constant after all. Furthermore, observations at high redshift of fully-formed galaxies with massive black holes at their centers by the James Webb Space Telescope, as well as black holes with unexpected properties observed by gravitational wave detectors LIGO-Virgo, are driving an in-depth revision of our assumptions in models of structure formation and the evolution of the Universe. I propose to explore two new paradigms to account for Dark Matter and Dark Energy, based on known physics without the need to introduce new particles in the Standard Model of Particle Physics. I will extend the primordial spectrum of fluctuations to small scales with new statistical properties to provide a viable Primordial Black Hole scenario for Dark Matter, and will include non-equilibrium thermodynamics in the expanding Universe, in the form of General Relativistic Entropic Acceleration, to explain Dark Energy. My proposal could provide a unified explanation for a plethora of interrelated multi-epoch, multi-scale, and multi-probe observations from present and future Gravitational Wave detectors, Large Scale Structure observatories, and Cosmic Microwave Background experiments. It emphasizes the need to develop new theoretical ideas hand-in-hand with observations to acquire a deeper understanding of our universe. If these ideas are correct, they will open a new window into the early universe and a new fundamental understanding of gravity in the late universe.
\end{abstract}

\maketitle

\section{Introduction}

Cosmology is a mature science in which the $\Lambda$ Cold Dark Matter ($\Lambda$CDM) model~\cite{Peebles}, with Gaussian initial conditions from cosmological inflation, provides the generally accepted description of our Universe. This paradigm was established thanks to a large number of cosmological observations from the last 30 years, which confirmed the accelerated expansion of the universe, the cold nature of dark matter, the amplitude and statistics of the primordial spectrum of perturbations, and the large-scale structure of the cosmic web. Despite these observational successes, the origin of its dark energy (DE) and dark matter (DM) components is still a mystery~\cite{Amendola, Scott}. No new particles, generally assumed to constitute the DM, have been found so far in high-energy colliders, astrophysical observations, or underground detectors. In the case of DE, the standard explanation for the accelerated expansion is the existence of a cosmological constant $\Lambda$, but with a value 120 orders of magnitude smaller than expected in Quantum Field Theory (QFT) and the Standard Model (SM) of Particle Physics. 

With the increasing precision of observations from ground and space telescopes, tensions and anomalies to the established paradigm begun to appear at small and large scales~\cite{RoyalSoc,Verde2019HubbleTension,DiValentino2021Intertwined,Bull2016FutureProbes}: too early galaxy formation, the origin of supermassive black holes (BH), the abundance of small-scale structures, the core-cusp problem in galaxies, cosmic dipoles, weak lensing anomalies, larger Integrated Sachs Wolfe (ISW) decrement in voids and the present Hubble rate tension. Moreover, the $\Lambda$CDM model has presently been challenged by at least three key observations: 1) the detection of gravitational waves (GW) emitted by binary black hole (BBH) mergers with black holes in the lower and upper mass gaps~\cite{GWTC3}, 2) the James Webb Space Telescope (JWST) detection of fully formed galaxies at very high redshifts~\cite{Carniani} ($z\sim14$) with massive BH at their centers, and 3) the recent results of DESI-BAO~\cite{DESI-BAO-DR1,DESI-BAO-DR2} and DES-SN~\cite{DES-SN-Y5,DES-SN-BAO} which, if confirmed, point to a non-constant $\Lambda$. All these new findings are challenging our ideas on the evolution and the formation of structure in the Universe.

My insight in the last few years has been that it is possible to explain DM and DE with known physics – General Relativity (GR), Thermodynamics, QFT in curved space-time, and the Standard Model of Particle Physics – but going beyond its usual applications, without having to resort to new particles beyond the SM nor introducing new degrees of freedom. The key ingredients are to extend the primordial spectrum of fluctuations to small scales with new statistical properties and to include non-equilibrium thermodynamics in the expanding Universe. 

In this review, I re-explore previous scenarios of primordial black holes~\cite{PBH} (PBH) as natural candidates for CDM and open new avenues like General Relativistic Entropic Acceleration~\cite{GREA} (GREA) to explain DE. As a concrete working hypothesis, I propose that CDM is entirely made of PBH, in a novel scenario in which the Higgs of the SM non-minimally coupled to gravity is the inflaton~\cite{CHI}, and where quantum diffusion generates curvature fluctuations on all scales, from the CMB anisotropies to PBH scales~\cite{QDiff}. The large curvature perturbations on small scales collapse upon reentry in the radiation era, generating PBH of different masses due to sudden pressure variations in the thermal evolution of the Universe~\cite{THM}. Supporting this scenario, some BBH events from GW detectors and the observations of the so-called Little Red Dots (LRD) by JWST~\cite{LRD1,LRD2} provide evidence for the widespread existence of unexpected black holes across different masses and redshifts, suggesting a non-astrophysical origin in compact PBH clusters~\cite{PBH-cluster2}. Additionally, we propose a radically new origin for DE, explained by the recently developed GREA theory~\cite{GREA-Basic}. GREA emerges from the growth of entropy~\cite{Jacobson1995ThermoGR,Padmanabhan2010EntropyGravity,Verlinde2011Entropic,Bousso2002Holographic,FischlerSusskind1998HolographicBound,Zhao2017DynamicalDE,Planck2020Params} of the boundary of our universe, our causal horizon~\cite{GREA-DE}. According to QFT in curved spacetime, such a boundary has intrinsic quantum entropy that grows over time. This growth induces a classical entropic force term in the Einstein equations, leading to an effective viscosity in the cosmic fluid that changes with time. This emergent viscosity gives an effective {\em negative} pressure in the energy-momentum tensor. While negligible in the distant past, this viscous pressure becomes the dominant component as the horizon expands, driving the current cosmic acceleration. Evidence supporting GREA includes its potential to resolve many of the cosmic tensions and explain the coincidence problem~\cite{GREA}.

The main challenge for these new proposals is to provide a unified explanation for a plethora of multi-epoch, multi-scale and multi-probe observations which are interrelated, since what we observe in the Universe today with optical and gravitational wave probes depends on what were the initial conditions during a period of inflation, together with its dynamical evolution in a matter/energy/entropy content with black holes and cosmic horizons. It is the right time to develop these theoretical ideas hand-in-hand with observations, in order to interpret the present and upcoming data and extract the key information that may lead to a deeper understanding of our Universe. With the planned next-generation experiments – in Large-Scale Structure (DESI, LSST, Euclid, Roman), Cosmic Microwave Background (Simons Observatory, LiteBird satellite), and Gravitational Waves (LVK, A+, Voyager, Einstein Telescope, Cosmic Explorer) –, my theories will be tested in the next 5 to 10 years. If confirmed, it will open up new scientific horizons and radically change our views of Dark Matter and Dark Energy.

\section{The scientific proposal}

The strategic objective of my proposal is to explore alternatives to $\Lambda$CDM in which all the DM is in the form of clustered PBH with an extended mass function, and DE is a consequence of the entropy growth of the causal horizon and the cosmic web. This program is divided into four specific objectives:

\begin{itemize}
    \item Theoretical developments in PBH: a) Obtain with quantum diffusion the precise curvature fluctuations from the CMB scale to small scales. b) Develop PBH cluster simulations to extract mass distributions and orbital characteristics of the BBH population. c) Construct N-body simulations of non-perturbative PNG tails.
    \item Searches of PBH: a) Gravitational Waves from BBH mergers. b) Multimessenger \& LSS. c) Microlensing surveys.
    \item Observational evidence/constraints on GREA: From LSS and CMB.
    \item Theoretical development of GREA: a) The cosmic web. b) Inflation and preheating.
\end{itemize}

\subsection{Theoretical developments in Primordial Black Holes}

For PBH to form in the radiation era, a large curvature gradient is needed to overcome the radiation pressure and produce gravitational collapse~\cite{PBH}. Such an enhanced gradient is more probable in cases where the distribution of primordial fluctuations from inflation has large non-Gaussian tails, as expected from quantum diffusion during inflation~\cite{QDiff,Sasaki2018PBHreview,YoungByrnes2015NG,Kawasaki2016PBHspikes,Byrnes2019PBHconstraints,Dalianis2021QCDPBH,Starobinsky1994Stochastic,Winitzki2000Diffusion}. In the context of the Thermal History of the Universe~\cite{THM}, with the known interactions of the Standard Model of particle physics, as the Universe expands, it goes through energy thresholds where large numbers of relativistic degrees of freedom disappear from the plasma (because they condense to form baryons and mesons, or annihilate with their antiparticles). This sudden change in the radiation pressure at the ElectroWeak, QCD, $\pi^+\pi^-$ and $e^+e^-$ scales exponentially increase the probability of collapse of the fluctuations that enter the horizon at that scale, giving rise to a very wide spectrum of PBH masses with distinctive peaks~\cite{THM} at planetary masses, stellar masses, IMBH and SMBH scales, providing the seeds for small scale structures in the universe~\cite{PBH-Hybrid}. This Thermal History Model (THM) evades all present microlensing and CMB constraints~\cite{PBH}. In order to achieve this objective, we propose the following tasks with their methodology:

a) Obtain with quantum diffusion the precise curvature fluctuations from the CMB scale to small scales. The process of PBH creation through gravitational collapse and the formation of early binaries (before recombination) depend on the amplitude and statistical properties of the primordial curvature fluctuations. We have recently shown that quantum diffusion during single field inflation can give rise to large exponential tails in the curvature perturbation distribution~\cite{QDiff}. This enhances not only the probability of collapse to form black holes in the radiation era, but also the probability above average of finding another black hole nearby, i.e. induces a primordial correlation function~\cite{QDiff-cluster} beyond Poisson clustering, which can then be used to estimate the size of clusters. In general, Primordial Non-Gaussianity (PNG) is non-perturbative and scale-dependent, so the usual parametrization in terms of a constant fNL is not enough, as we demonstrated~\cite{ElGordo}, and we have to go beyond this first-order approximation and include a fully non-perturbative approach. In this project, we will perform a full analysis of quantum diffusion in a concrete model like Critical Higgs Inflation~\cite{CHI}, which will give us the probability distribution function of curvature fluctuations on all scales. To see the effect of these PNG exponential tails on the probability of PBH formation and clustering, we also need to take into account the critical collapse threshold as a function of the thermal history of the universe~\cite{PBH}. This will provide the initial mass and spatial distribution at formation as input for the next objectives b) and c).

b) Develop PBH cluster simulations to extract mass distributions and orbital characteristics of the BBH population. The parameters and merger rates of BBH events detected by GW interferometers depend not only on their initial mass and spatial distribution at formation, but also on the dynamical environments in which the binaries form and evolve. In the case of PBH, we expect the binaries to be formed in dense clusters before or after recombination, even with simple Poisson clustering~\cite{PBH}, in which the binaries form and disrupt stochastically, with a non-negligible fraction of them being kicked off the cluster and eventually merging in the outer halos of galaxies~\cite{PBH-cluster1}. I will study these complex dynamics in detail with different initial conditions and a very large number (of order a few tens of thousands) of PBH using a Nbody6++ code~\cite{PBH-cluster2} and the new PeTar code~\cite{PeTar}, and compare with the predictions from binary BH formation through astrophysical channels. The main observables to extract from these simulations are the mass distribution of binaries, the mass ratio of components, the initial separation, eccentricity and hardness of the binaries, and the spin distribution of both components. Most of these distributions are at present chosen “ad-hoc” or, in the case of the binary formation rate in PBH clusters, calculated with an approximate formula obtained from unrealistic and very limited simulations, which gives a great degree of uncertainty in the PBH predictions and constraints~\cite{LISA-PBH}. It is thus of crucial importance to obtain realistic predictions for PBH binary formation rates, not only for the analysis of the LVK data but also, if the PBH population is confirmed, for the development of the right physics program of future GW observatories (LISA, Einstein Telescope, Cosmic Explorer, etc.).

c) Construct N-body simulations of non-perturbative PNG tails. Primordial Non-Gaussianities could also be responsible for an increase in size and abundance of small-scale structures and the seeds of galaxies at high redshift, as observed by the JWST~\cite{QDiff-cluster}. In this project, I intend to study, with dedicated N-body simulations, the effect of PNG with exponential tails in the matter fluctuation spectrum and follow the non-linear gravitational collapse in a CDM-only simulation to see whether the PNG leaves imprints on large-scale structures like supervoids and superclusters, following the techniques developed by my group in PNG-UNITsims~\cite{UnitSim} for DES and DESI. This expertise in DES will also be used to measure PNG within Euclid and LSST.

\subsection{Searches of Primordial Black Holes and constraints}

a) PBH searches with GW. The BBH observation plan of the LVK collaboration is suitable for PBH searches in three regions of interest: sub-solar mass (SSM)~\cite{Zevin2021Multichannel,Belczynski2020PISN,Magee2022Subsolar,HallGow2020SubsolarTheory,KohriTerada2018SGWB,Domenech2021ReviewInducedGW}, lower mass gap (LMG) (mass between 2.5 and 5 solar masses), and pair-instability supernova mass gap (PISN) (mass in the 60-120 solar mass range). In the last two regions, several candidates have been found that challenge astrophysical interpretation. SSM black holes, with no standard astrophysical production channels, are natural PBH candidates. I was paper manager of the SSM searches in run O3b of LVK~\cite{LVK-SSM}, where my group has pioneered the first in-depth analysis of the most significant SSM candidates found in run O2 and O3 data~\cite{SSM1, SSM2} and developed new faster ROQ tools~\cite{ROQ} needed for the parameter estimation of these long-duration events with small chirp masses and mass ratios. We are actively contributing to the O4 SSM analysis, and I also plan to study correlations between pipelines and signal coherent tests among the three LVK detectors on simulated SSM signals and background triggers, in order to properly assess the significance of the SSM candidates. In the lower mass gap, there are two confirmed compact objects~\cite{LMG}, both most likely black holes, due to their mass and lack of tidal deformation. No known supernova core-collapse models can generate these types of objects. In particular, GW190814~\cite{GW190814}, with its secondary BH in the LMG, highly asymmetric masses and zero spin, challenges all existing astrophysical models~\cite{GW190814}, while events like this one are expected in the THM~\cite{THM}. I plan to promote within LVK and participate in the analysis of a specific search for low-mass-ratio binaries within the O4 and O5 LVK runs. Moreover, GW190521~\cite{UMG} is a confirmed BBH event with both components in the PISN mass gap~\cite{UMG}. They could come from a previous BH merger, but the detailed analysis of the BH progenitor spins and natal kicks makes it unlikely to be formed in standard astrophysical environments~\cite{astro}. On the other hand, the THM predicts a peak (in fact a "shoulder") in the PISN mass gap due to pion annihilation~\cite{pion}, while providing initial dense PBH clusters for binary formation. Once we have enough statistics in the merger rate, mass, spin, and redshift distributions and their correlations, I will use the results of our cluster simulations together with LVK population analysis tools to determine if there are two populations of black holes (astrophysical and primordial) in the LVK Transient Catalogs. Moreover, if a primordial population is found with peaks as predicted by the thermal history of the Universe, their measurement can provide feedback on the underlying spectrum of fluctuations at small scales and on the physics of the early universe.

Large amplitude fluctuations needed for sub-solar mass PBH formation source a Stochastic GW Background from second order perturbations in the LISA band~\cite{Bartolo}. I am leading two PBH projects within the LISA Cosmology Working Group to explore the signatures of PBH on mHz-frequencies~\cite{LISA-PBH}. In particular, we are developing the tools (with an end-to-end code) to predict, search, and analyse/constrain all the possible signatures from PBH. 

b) Multimessenger and Large Scale Structure. The era of Multimessenger astronomy started with the kilonova GW170817~\cite{kilonova}, a binary neutron star (BNS) merger that allowed to measure simultaneously its cosmological distance with GW and its redshift with its EM counterpart~\cite{EMcounterpart}, providing the first “standard siren”. I was in the DES-GW team for the discovery paper of the EM source in NGC4993 and also contributed to the analysis of the peculiar velocity systematics in the precise measurement of the present rate of expansion~\cite{Nature}. For the moment, we have only detected this clear BNS event at a relatively nearby distance of 40 Mpc, but by the end of LVK run O5, we expect dozens of events at distances up to 300 Mpc, in whose analysis I will contribute from both the GW side in the LVK Cosmology group and in the DECam EM Counterpart follow-up team. We aim to determine H0 with error bars comparable to those of the local SN-Ia, and thus address the so-called Hubble tension with different systematics. Furthermore, well-localized BBH mergers can be used as “dark sirens”~\cite{Palmese2020DarkSiren,MukherjeeWandelt2018DarkSiren} to statistically locate the host galaxy (or cluster of galaxies) by correlating with known galaxy catalogues~\cite{Schutz}. For the expected rates of events in the next observation runs, a sufficiently large number of these sources could be used to statistically determine both the matter distribution and the rate of expansion of the universe, as we demonstrated~\cite{DarkSiren}. The spatial distribution of dark sirens can give us insight into the nature of DM through the cross-correlation with the LSS distribution. For dark sirens of astrophysical origin that come from the inner parts of galaxies, one expects the correlation to be unbiased, while if they come from PBH in the outer halos of galaxies, where they form the dark matter, it should have bias one~\cite{bias,Bird2016PBHLSS,Hutsi2021PBHclustering}. Finally, the inferred distributions from my PBH cluster simulations will help to clarify if LVK events classified as BNS or NSBH solely because of their mass (without EM counterpart) are instead BBH from the QCD peak in the THM. This has important consequences for the estimate of BNS and NSBH merger rates of events.

c) Microlensing search for PBH of solar-mass range. Microlensing surveys towards the LMC have put stringent constraints on the abundance of single-mass and uniformly distributed PBH in the halo of our galaxy. However, these constraints are significantly relaxed if PBH are clustered and with a wide mass distribution as in the Thermal History Model~\cite{PBH}, still allowing PBH to be 100\% of the DM. The Gaia satellite has found 363 new microlensing events~\cite{Gaia-MLE} and will improve our knowledge of the galaxy matter distribution, to determine the edge of the DM halo~\cite{Halo-MW} and correctly estimate the optical depth to the LMC. In fact, my recent reanalysis of the MACHO constraints with Gaia DR3 data~\cite{Gaia-DR3} showed that there are assumptions made by microlensing surveys about the extent of the halo and the rotation curve of our galaxy, that can change significantly the conclusions on the abundance of PBH, and reinterpret the observed microlensing events as due to compact objects rather than stars~\cite{Hawkins}. For this reason, we have proposed, within the Legacy Survey of Space and Time (LSST) Consortium in the Vera Rubin Observatory, a new microlensing survey with LSST and the Nancy Roman space telescope (NRST)~\cite{Abrams}. This combined photometric (LSST) and astrometric (NRST) survey will allow us to break the degeneracies between mass and distance in the Einstein radius and thus half-crossing times~\cite{Fardeen}. I further proposed that we use spectroscopy of the source in order to distinguish the type of dark compact object that acted as lens. This may allow us to eventually detect the QCD peak of the THM in the solar-mass range, with a sub-solar mass tail that would be proof of its primordial origin. 

\subsection{Observational constraints on General Relativistic Entropic Acceleration}

The effects of GREA are expected on both large and small scales~\cite{GREA}, from the horizon to the cosmic web, the formation of SMBH, galaxies, clusters, and supervoids. The predictive power of GREA on large scales is a consequence of its simplicity, described by a single parameter relating the size of the horizon to the curvature scale~\cite{GREA-DE}. This makes GREA a very attractive alternative to $\Lambda$CDM to explore, not only because it is based on first principles~\cite{GREA-Basic}, but also because its large-scale prediction can be falsified given the expected precision of next-generation cosmological observations. In case of being confirmed, it will give a first-principles explanation of both the present acceleration and the extremely small value of the effective cosmological constant. 

a) Constraints from LSS and CMB. The future precision in the measurement of multiple observables – the location of the coasting point, the rate of expansion today, the full redshift evolution of the “effective” equation of state of dark energy, w(z), the luminosity and angular diameter distances from SN-Ia~\cite{DES-SN-BAO} and BAO~\cite{DESI-BAO-DR2} observations, the age of the universe, as well as the effects on Redshift Space Distortions (RSD), Integrated Sachs-Wolfe (ISW), the BAO scale evolution and the full spectrum of CMB anisotropies – requires a corresponding precision in the calculation of the evolution of the background and the linear perturbations in GREA, see ref.~\cite{GREA}. I will use Boltzmann solvers like CLASS~\cite{CLASS}, incorporating the proper time evolution of the scale factor and the density contrast, to obtain the above observables within GREA, which will then be used to run Monte Carlo simulations like MontePython~\cite{MontePython} for estimation of the likelihood and determine the Bayes ratios between the $\Lambda$CDM model and the GREA hypothesis. A first estimate using the DESI-DR2 BAO~\cite{DESI-BAO-DR2} has already been used to put contraints on the effective parameter $\alpha$ of GREA~\cite{GREA-DESI}.

b) Consequences for small scales. I also plan to study the consequences and potential observability of GREA on the dynamics and evolution of small-scale perturbations. GREA from mass accretion onto black holes has potential signatures, e.g., in the far past, the rapid growth of SMBH could provide a substantial local acceleration to the universe that may have affected early structure formation~\cite{GREA-SMBH}. In an ongoing DESI project, I will explore the rate of growth of SMBH in the centre of QSO to determine the redshift-dependence of this GREA effect. In the late universe, BBH mergers produce a sudden increase in entropy of the BH and the emission of gravitational waves that carry away a fraction of the entropy, possibly affecting the merger kicks. I plan to study this effect on the GW signals from large signal-to-noise-ratio BBH mergers.

\subsection{Theoretical developments of General Relativistic Entropic Acceleration}

a) The cosmic web. The creation of order that characterizes the large-scale distribution of matter in the cosmic web, like superclusters of galaxies, must come together with the corresponding generation of entropy on those scales. According to GREA, this flow of entropy should be responsible for an entropic force that shapes large-scale structures like supervoids, making them deeper than would be expected from the overall accelerated expansion of the cosmic horizon~\cite{GREA}. This extra entropic force could be detected via the Integrated Sachs-Wolfe (ISW) effect along the line of sight of deep and large voids in our vicinity, since it will tend to generate an extra decrement in the temperature of the CMB beyond that which can be accounted for by the global accelerated expansion of the universe and will be specific to the particular location of the void~\cite{void}. I intend to develop new N-body simulations using GADGET-4 based on parallel particle mesh to describe the non-linear gravitational collapse in the presence of inhomogeneous GREA, assuming initial conditions in the CMB, to see how GREA induces deviations with respect to $\Lambda$CDM at small and large scales.

b) Inflation and preheating. Particle production at (p)reheating is a quantum process that generates a fantastic amount of entropy in a fraction of a second before thermalization~\cite{KLS}. GREA will induce a short burst of acceleration of a few extra e-folds soon after inflation. A possible consequence is the generation of a burst of metric fluctuations, both scalar and tensor, with an increase in the amplitude of GW on small scales that could be seen in high-frequency GW detectors~\cite{UHFGW}. I will explore the corresponding particle physics phenomenology with the code CosmoLattice~\cite{CosmoLattice}. Cosmological inflation is usually described with an effective field theory, characterized by a scalar field with a concrete potential~\cite{Linde}, whose fluctuations generate the curvature perturbations responsible for CMB anisotropies and Large Scale Structures in the late universe. In this project, I will explore a radically new approach to primordial inflation based on GREA. At the Planck scale, space-time is subject to violent quantum metric fluctuations that collapse to form BH, which evaporates during a tiny fraction of a second~\cite{Hawking}. In this process, BHs emit particles, radiating away the entropy in the horizon as they shrink in mass and size. I will explore how this increase of entropy in the vicinity of the BHs could drive an entropic acceleration that lasts until these particles thermalize via Standard Model interactions at around $10^{15}$ GeV. In this scenario, GREA would act like the trigger (Big Bang) that started the expansion of the universe, and may create a rich arena for new phenomenological signatures that could be tested with CMB experiments like the LiteBird satellite and the Simons Observatory. Moreover, it may be one of our few windows into the Principle of Holography and Quantum Gravity~\cite{GREA-holo}.

\section{Discussion and Conclusions}

The undertaking of a scientific program of this caliber inevitably involves risks. The hypotheses are based on well-founded physics (the primordial origin of CMB fluctuations, the thermal history of the universe, and the interactions between particles) to explain fundamental problems with new approaches. The risk is whether the proposed scenarios are enough to explain such a large and interconnected phenomenology.  I have presented a detailed strategy to test my hypotheses with the right combination of theoretical developments (to refine the predictions), multi-probe experimental methods (to prevent confirmation bias), and multiscale observations (to prove its consistency), which substantially mitigates the potential risks. Since my hypotheses imply a high degree of coherence over many different observations, we can learn from observables that do not match predictions, and this may point to the specific components of the theoretical scenarios that need to be refined, ensuring advances in the field. With a track record of groundbreaking theoretical proposals in the last decades (in inflation, preheating, non-Gaussianities, large voids, etc.) and my knowledge of the cosmological observations’ caveats and intricacies, due to my extensive experience in LSS and GW collaborations (DES, PAU, DESI, LVK, Euclid, LISA, ET, etc.), I am optimally positioned to achieve the objectives of this project. 

This project aims to push the boundaries of cosmology by exploring the ideas of quantum diffusion and large curvature fluctuations from inflation to explain dark matter, and the Bekenstein entropy with Gibbons-Hawking boundary term to explain dark energy. I will lead this research since I was one of the first to propose PBH as DM from inflationary fluctuations in the 1990’s and led the renaissance of PBH phenomenology in the last decade. Recently, I have also proposed GREA as an inevitable consequence of generally-covariant out-of-equilibrium thermodynamics, and the origin of an effective entropic force associated with the quantum-gravitational degrees of freedom of horizons. 

The scenario described above has very concrete predictions which will be tested in the next 5 to 10 years, thanks to the new generation of cosmological experiments and GW detectors that are starting now or already approved. Moreover, the scientific program proposed will provide essential input for the technological design and physics science case of next-generation experiments in LSS, CMB, and GW. If any or both of these ideas, PBH and GREA, are correct, we open a new window into the early universe and a new fundamental understanding of quantum gravity in the late universe. It will also provide a vision into our place in the cosmos and enhance the visibility of European missions like LISA and Euclid. The overall impact of this project can be transformative, with radically new paradigms that could drive theory and experiment for many decades. 

\vspace{6pt} 

\section{Acknowledgements}

The author acknowledges support from the Spanish Research Project PID2024-159420NB-C43 [MICINN-FEDER], and the Centro de Excelencia Severo Ochoa Program CEX2020-001007-S at IFT. 

\bibliography{DMDE}

\end{document}